\documentclass[twocolumn,showpacs,amsmath,amssymb,amsfonts,nofootinbib,prl]{revtex4-1}
\usepackage{graphicx}
\usepackage{dcolumn}
\usepackage{bm}
\usepackage{ulem}
\usepackage{overpic}

\def\beq{\begin{equation}}
\def\eeq{\end{equation}}

\usepackage{soul}
\usepackage{cancel}

\usepackage[usenames]{color}

\begin{document}
\input{epsf}

\title{Jet-parton inelastic interaction beyond eikonal approximation}

\author{Raktim Abir}
\affiliation{Theory Division, Saha Institute of Nuclear Physics,  
1/AF Bidhannagar, Kolkata 700064, India.}

\begin{abstract}
Most of the models of jet quenching  generally assumes that a jet always travels in a straight eikonal path, which is indeed true for 
sufficiently hard jet but may not be a good one for moderate and low momentum jet. In this article an attempt has been made to 
relax part of this approximation for $2\rightarrow 3$ processes and found a (15-20)$\%$ suppression in the differential cross-section 
for moderately hard jets because of the noneikonal effects. 
In particular, for the process $qq' \rightarrow qq'g$ in the centre of momentum frame the scattering with an
angle wider than $\pm0.52\pi$ is literally forbidden unlike the process $gg\rightarrow ggg$ that allows 
an angular range $\pm \pi$. This may have consequence on the suppression of hadronic spectra at low transverse momenta.   
\end{abstract}


\date{\today}
\maketitle

\section{Introduction}

Constituent {\it quark number scaling} of elliptic flow and {\it jet quenching} are supposed to be the most prominent signatures that favour 
the partonic degrees 
of freedom in the deconfined QCD matter. Overwhelming evidences of both signatures coming from dedicated heavy ion experiments $viz.,$ 
STAR and PHENIX{\it@} RHIC BNL~\cite{Adams:2005dq,Adcox:2004mh}, ALICE {\it@} LHC CERN~\cite{Aamodt:2010jd,Aamodt:2010pa} established the fact that, 
the primordial hot-soup of nuclear matter, produced in those experiments, indeed contain {\it partonic degrees of freedom},
before freeze out to hadrons in later stage, instead of {\it hadronic degrees of freedom} throughout. Observation of strong suppression of inclusive yields of high momentum hadrons and semi-inclusive rate 
of azimuthal back-to-back high momentum hadron pairs relative to  {\it p-p} collisions, are expectations from jet quenching. Both of them are 
extensively explored in collisions of {\it Au-Au} nuclei at $\sqrt{s}=200$ A GeV in RHIC. ALICE, the dedicated heavy Ion collider experiment at 
CERN, seems to appear as a factory of Jets. Evidence for jet quenching has also been observed recently in {\it Pb-Pb} collision at 
ALICE \cite{Aamodt:2010jd}.

There are a few well known models in the 
literature~\cite{Salgado:2003gb,Wicks:2005gt,Wang:2001ifa,Jeon:2003gi,Mustafa:1997pm,Mustafa:2003vh,Qin:2007rn,Qin:2009gw,Abir:2012pu,Abir:2011jb,Zapp:2012ak} that aim to 
quantify energy loss (mainly radiative) and jet quenching phenomena within perturbative QCD. Most of these formalisms  
for energy loss of a high momentum parton through gluon radiations, have some common technical approximations in order to
make the calculation simpler. These approximations are implemented 
both at the level of {\it single emission kernel} calculations and at {\it multiple gluon emission estimation} schemes. Main kinematic approximations at the 
level of {\it single emission kernel}, are listed below (also see  Refs.~\cite{Armesto:2011ht,Mehtar-Tani:2013pia,Renk:2011aa} for a comprehensive discussion):

\begin{itemize}
\item {\it Eikonal parton trajectory~I~:~} 
 Leading parton is having energy $E$ ($i.e.$ $p_z = E$ and $p_\bot=0$, by definition to start with)  is much larger 
 than the transverse momentum exchanged gluon $q_{\perp}$ with the medium, 
 $E \gg q_{\perp}$, so that it does not give sufficient {\it transverse kick} to deflect the parton from straight trajectory along $z$ axis. 
To relax this approximation it is therefore important to keep track of the terms of ${\cal O}(q_\bot/E)$ in the formalism. 
 
\item {\it Eikonal parton trajectory~II~:~}  Energy of the leading parton is sufficiently high, $E \gg k_{\perp}$ ($k_{\perp}$ being transverse 
component of the emitted gluon) so that it does not get enough {\it transverse kick} also from emitted gluon. This ensures that the leading parton is 
in eikonal trajectory. {\it However it does not fix any definite direction for gluon emission, which requires comparison of $k_{\perp}$ with 
longitudinal component $k_{z}$ or with energy $\omega$ of the emitted gluon. Therefore, it is important not to neglect terms of ${\cal O}(k_\bot/E)$ 
in the formalism to relax this approximation in the jet studies. }    

\item {\it Soft gluon emission~:~} One often uses the additional approximation that the gluon 
energy is much smaller than the leading parton energy $\omega \ll E$. When $x$ is the fraction of {\it energy} carried out by
the  emitted gluon, $i.e.$, $x=w/E$, this approximation ensures that $x \rightarrow 0$. When $x$ is typical light cone variable, defined 
by the fraction of light cone $(+)ve$ momentum carried out by the emitted gluon, $i.e.$, $x=k^+/p^+= (k_\bot/\sqrt{s})e^{\eta}$, this approximation 
ensures $x\rightarrow 0$, only {\it in the mid rapidity and backward rapidity regions $(-\infty \geq \eta \geq 0)$ but not in the forward 
regions where $\eta$ could be a large positive number}. 

\item {\it Small angle/collinear gluon emission~:~} Energy of the emitted gluon $\omega$ is much larger 
than its transverse momentum $k_\bot$, $\omega \gg k_\bot$ and $\omega \simeq k_{z}$. 
For any $2\rightarrow 3$ process, without loss of generality, one can take $k_\bot = \omega \sin\theta_g$ 
and $k_{z} = \omega \cos\theta_g$, where $\theta_g$ being the angle between direction of propagation of 
leading parton and direction of emitted gluon. This particular approximation therefore implies $\theta_g \simeq 0$.
\end{itemize}

At this point it is worth mentioning that {\it soft gluon emission} approximation is a broader approximation as it automatically encompasses 
the {\it eikonal parton trajectory~II} approximation, because energy $w$ should 
always be more than the transverse momentum $k_\bot$ for a massless emitted gluons.

In this article we make an effort to relax the {\it eikonal parton trajectory~I} approximation in some extent for the radiative/inelastic 
process $2\rightarrow 3$. Investigation of all the matrix elements in ${\cal{O}}(\alpha_s^3)$ for  $2\rightarrow 3$ radiative processes 
have been done keeping terms up to $\cal{O}$ $(t/s)$. The  first order in eikonal expansion, {\it i.e.}, $\cal{O}$ $(t/s)$, is termed
here as {\it noneikonal}, since  the calculation is performed in Feynman gauge with Mandelstum variables instead of most extensively used light 
cone gauge with light cone variables\footnote{After connecting  $t$ to $q_{\perp}$ and $s$ to $E$ in the centre of momentum frame, term of $\cal{O}$ $(t/s)$ 
ensures the relaxation of the approximation $E \gg q_{\perp}$ ({\it eikonal parton trajectory~I}).}. We have neglected 
terms of ${\cal{O}}(t^2/s^2)$ and higher orders. Throughout our study we also do not use any approximation related to {\it small angle/collinear gluon emission}. 
Nevertheless, we have used {\it soft gluon emission approximation} which automatically include {\it Eikonal parton trajectory~II} assumption. 

Here we relaxed the kinematic constrains associated with eikonal propagation.
Nevertheless (non)eikonal propagation is directly related to
space-time tracks that the partons take when going through the medium.
While an eikonal path is a straight line, a non-eikonal path is
generically any trajectory that have deviation from straight line and obviously always somewhat longer. This
longer path, taken by the parton may matter, in the soft sector when the problem is embedded 
into a hydrodynamically evolving density distribution \cite{Renk:2011gj}. 
We note that there is transport model, although uses the cross sections which 
are kind of eikonal, takes all  the trajectories into account, so particles 
do not need to move on straight lines \cite{Fochler:2010wn}.

\section{ Inelastic Quark-Quark Scattering ($qq'\rightarrow qq'g$)}
\begin{figure}[t]
\centering
\begin{overpic}[width=0.6\linewidth]{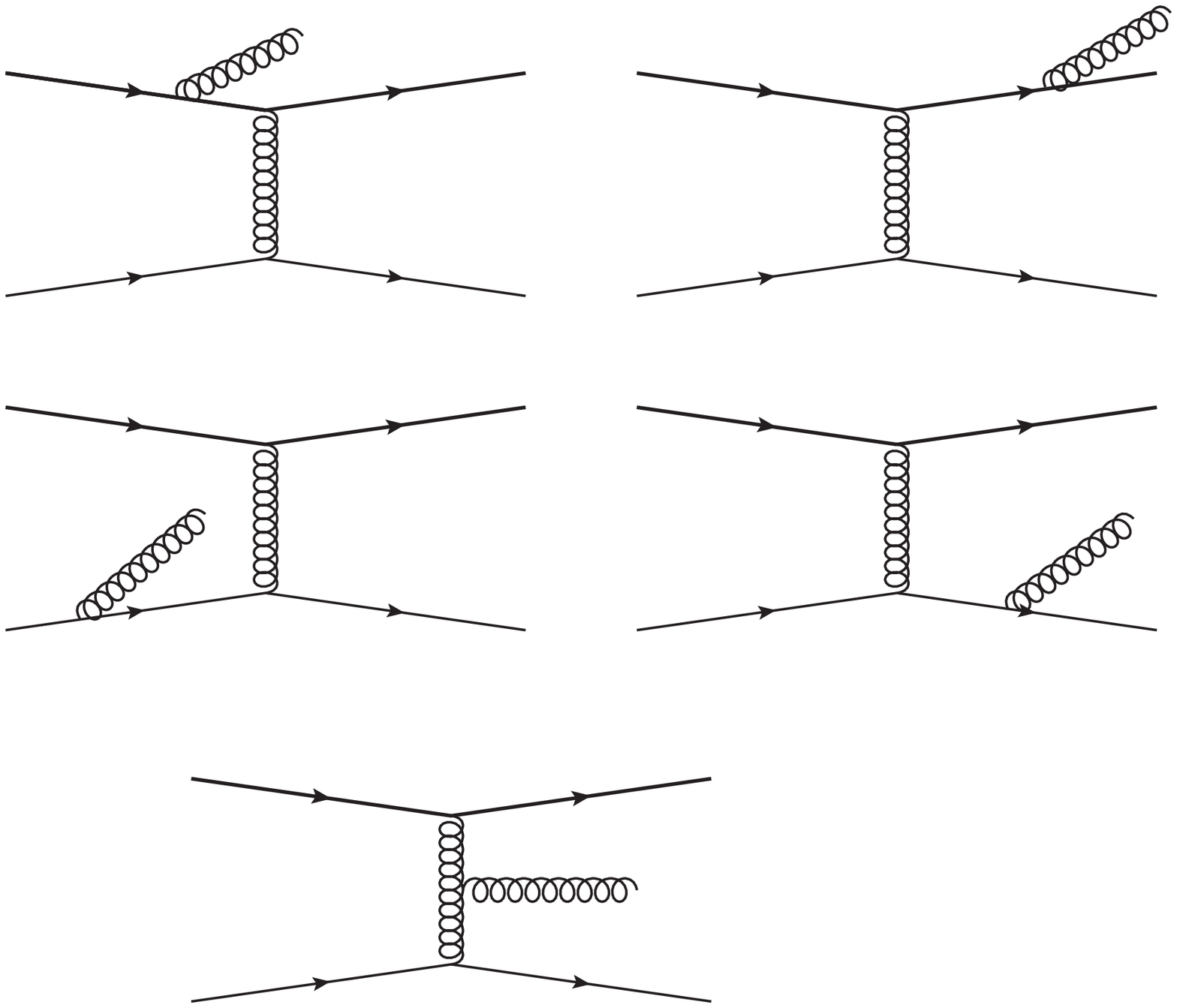}
\put(-8,67){(1)}
\put(-8,38){(2)}
\put(51,67){(3)}
\put(51,38){(4)}
\put(25,10){(0)}
\end{overpic}
\caption{Five tree level Feynman diagrams for the process $qq'\rightarrow qq'g$. 
}
\label{diagram}
\end{figure}

\begin{figure}[t]
\includegraphics[width=0.8\columnwidth]{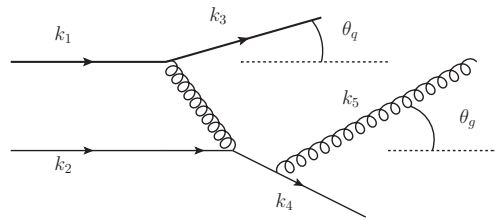}
\vspace*{-0.0in}
\caption{Non-zero angular deviation from eikonal trajectory. Angle between incoming and outgoing momentum of 
projectile is $\theta_q$ and direction of emition of gluon with that of incoming momentum of projectile parton is $\theta_g$.}
\label{fig1}
\end{figure}

The process $qq'\rightarrow qq'g$ (prime denotes different quark flavour) in  ${\cal{O}}(\alpha_s^3)$ 
appears in five $t$ channel Feynman diagrams, which are shown in the Fig.~\ref{diagram} (see also Fig.~\ref{fig1} for
other details). Note that $k_1$ and $k_2$ are momenta of the different quark flavours in the entrance channel
whereas  $k_3$ and $k_4$ are those for exit channel and $k_5$ is that of the emitted gluon. Scattering angle between ${\bf k_1}$ and ${\bf k_3}$ is $\theta_q$ whereas $\theta_g$ is the angle between direction of emission of soft gluon $\bf \hat k_5$ and direction of {\it incoming} projectile quark $\bf \hat k_1$. 
We now define the relevant Mandelstum variables for this $2\rightarrow3$ process as
\begin{eqnarray}
&&\hspace{-0.3cm}s=(k_1+k_2)^2\,,\hspace{0.3cm} 
s^\prime=(k_3+k_4)^2, \nonumber \\
&&\hspace{-0.3cm}u=(k_1-k_4)^2\,,\hspace{0.3cm} 
u^\prime=(k_2-k_3)^2, \nonumber \\
&&\hspace{-0.3cm}t=(k_1-k_3)^2\,,\hspace{0.3cm} 
t^\prime=(k_2-k_4)^2, \label{eq1} 
\end{eqnarray}
with
\begin{eqnarray}
 &&\hspace{-0.4cm} s+t+u+s'+t'+u'= 0 \ .
\label{eq2}
\end{eqnarray}

When the emitted gluon is soft ($k_5\rightarrow 0$) compare to other external legs, one can assume :
$t^\prime \rightarrow t, \ \ s^\prime \rightarrow s, \ \ 
u^\prime \rightarrow u$ and we can express the transverse component 
of the momentum of the emitted gluon in the centre of momentum frame of 
$k_1$ and $k_2$ as
\begin{eqnarray}
k_\perp^2 &=& \frac{4(k_1\cdot k_5)(k_2\cdot k_5)}{s}   
\nonumber \\
&=&
\frac{(s+t+u)(s+u'+t')}{s}\nonumber \\
&=&\frac{(s+t+u)^2}{s} \ . 
\label{eq3}
\end{eqnarray}

%

The hierarchy among various scale of momentum, employed in the present work is stated as
\begin{equation}
\sqrt{s},E > \sqrt{|t|}, q_\bot \gg w \geq k_\bot \geq m_d \, ,
\end{equation}
where $m_d$ is the Debye screening mass acts as an infrared cut-off.
We note that the above hierarchy relaxes the aproximations  $\sqrt{s},E \gg \sqrt{|t|}, q_\bot$ 
({\it Eikonal parton trajectories}~I ) and $w \gg k_\bot$ 
({\it small angle/collinear gluon emission }) in some extent.

\subsection{Matrix Elements, Amplitude and Cross-section}
The gauge invariant amplitude summed over all the final states 
and averaged over initial states for the process, 
$qq'\rightarrow qq'g$, is 
\begin{equation}
\left | {\cal M}_{qq'\rightarrow qq'g}\right |^2 = \sum_{i\ge j} 
{\cal M}_{ij}^{ 2} \ , \label{eq5}
\end{equation}
where $i$ and $j$ run from $0$ to $4$. We note that the index $0$ 
represents the diagram where the soft gluon emits from the exchanged gluon line whereas
the indices $m=1,2,3,4$ represent the diagrams where emission of the soft gluons are being from external
fermion lines having momenta $k_m$ (see Fig.~\ref{fig1}). Equation (\ref{eq5}) contains  
total fifteen terms in which there are five self interfering `genuine amplitudes' for $i=j$  
and ten cross interfering `interference amplitudes' associated with  gluon emissions involving 
two diagrams for $i\ne j$. Here we have extensively used the FORM, REDUCE and 
CalcHep Programs~\cite{Belyaev:2012qa}. Results are given below  up to 
${\cal O}(1/k_\bot^2)$ and ${\cal O}(t/s)$, for soft gluon emission. 

\subsubsection{Genuine amplitudes~}
By genuine amplitudes we are referring those are coming from each diagram that interfers with itself. 
In the Feynman gauge~\footnote{In  light cone gauge 
amplitudes coming from diagrams that involve gluon emission from target partons can only be neglected, others  are not.} 
all of them vanish within {\it soft gluon emission} approximations and in ${\cal O}(1/k_\bot^2 )$:
\begin{eqnarray}
{\cal M_{\rm 11}^{\rm 2}}&=&{\cal M_{\rm 33}^{\rm 2}} = 0 ; \, \,
{\cal M_{\rm 22}^{\rm 2}}={\cal M_{\rm 44}^{\rm 2}} = 0;  \, \,
{\cal M_{\rm 00}^{\rm 2}}= 0.
\end{eqnarray}
However, they may contribute in ${\cal O}(1)$, ${\cal O}(k_\bot^2)$, ${\cal O}(k_\bot^4 )$ etc, and in ${\cal{O}}(t^2/s^2)$ and higher orders.
All of them can safely be neglected in the soft emission limit as our aim is to go beyond the {\it eikonal approximation-I}, $E \gg q_{\perp}$.  

\subsubsection{Interference amplitudes~}
Within the approximations employed above  the amplitudes corresponding to the matrix elements of $({\rm 1} \bf \otimes {\rm 4})$ 
and  $({\rm 2} \bf \otimes {\rm 3})$ are identical in the leading order (${\cal O}(1/k_\bot^2)$) as well in ${\cal O}(t/s)$ and 
given as 
\begin{eqnarray}
{\cal M_{\rm 14}^{\rm 2}}&=&{\cal M_{\rm 23}^{\rm 2}}=\frac{7}{8}
\frac{128}{27}g^6\frac{s^2}{t^2}\frac{1}{k_\bot^2}\left[1+2\frac{t}{s}
\right],  \nonumber
\end{eqnarray}
where ${\cal O}(t/s)$ is purely noneikonal ({\it i.e.}, the first order in eikonal approximation) in nature as noted earlier. 
Also the amplitudes for $({\rm 1} \bf \otimes {\rm 2})$ and  $({\rm 3} \bf \otimes {\rm 4})$ are identical and 
the contribution is obtained in ${\cal O}(1/k_\bot^2)$ and ${\cal O}(t/s)$ as
\begin{eqnarray}
{\cal M_{\rm 12}^{\rm 2}}&=&{\cal M_{\rm 34}^{\rm 2}}=\frac{1}{4}
\frac{128}{27}g^6\frac{s^2}{t^2}\frac{1}{k_\bot^2}\left[1+\frac{t}{s}\right]. \nonumber 
\end{eqnarray}
Both $({\rm 1} \bf \otimes {\rm 3})$ and  $({\rm 2} \bf \otimes {\rm 4})$ are also identical within 
the employed hierarchy. However, they do not contribute in leading order but only in ${\cal O}(t/s)$. 
Hence, {\it in Feynman gauge} the contribution from the interference between {\it initial state } and {\it final state} radiations,
is exclusively noneikonal in nature and given as 
\begin{eqnarray}
{\cal M_{\rm 13}^{\rm 2}} &=& {\cal M_{\rm 24}^{\rm 2}} = \frac{1}{4}\frac{128}{27}g^6
\frac{s^2}{t^2}
\frac{1}{k_\bot^2}\left[\frac{1}{2}\frac{t}{s}\right] , \nonumber 
\end{eqnarray}

Finally any diagram interfering with  ${\rm 0}$, {\it i.e.,} $({\rm 0} \bf \otimes {\it l})$ with $l=1,2,3,4$, does not contribute 
in ${\cal O}(1/k_\bot^2)$ but contributes in $(1/k_\bot \sqrt{t})$ and higher orders. In the limit $|\sqrt{t}|\sim q_\bot\gg w$, amplitudes 
of ${\cal O}(1/k_\bot \sqrt{t})$ are subleading, in comparison to ${\cal O}(1/k_\bot^2 )$. Therefore, all of these  
 (${\cal M_{\rm 10}^{\rm 2}}$, ${\cal M_{\rm 20}^{\rm 2}}$,
 ${\cal M_{\rm 30}^{\rm 2}}$ and ${\cal M_{\rm 40}^{\rm 2}}$)
do not contribute within the approximation used in this work. 

Here we note that color coherence pattern between initial and final state radiation in the presence of a QCD medium derived in light cone gauge shows angular distribution of the induced gluon spectrum is broadly modified when one includes interference terms \cite{Armesto:2012qa,MehtarTani:2012cy}. In Feynman gauge within our hierarchy only interferance terms are contributing to the amplitude.

The gauge invariant amplitude for the process, $qq'\rightarrow qq'g$, can now
be obtained by summing all the subamplitudes as
\begin{eqnarray}
\left| {\cal M}_{qq' \rightarrow qq'g} \right|^2 
\!\!\!&=&\!\!12g^{2}\left| {\cal M}_{qq' \rightarrow qq'} \right|^2_{eknl} 
\frac{1}{k_\perp^{2}}\left(1+\frac{17}{9}\frac{t}{s}\right), 
\label{eq101} 
\end{eqnarray}
where the two body amplitude is given as
\begin{eqnarray}
\left|{\cal M}_{qq^\prime\rightarrow qq^\prime}\right|^2_{eknl} = \frac{8}{9}g^4\frac{s^2}{t^2}. 
\label{eq7}
\end{eqnarray}

The three-body amplitude in (\ref{eq101}) for the inelastic process, $qq'\rightarrow qq'g$, contains
the two-body amplitude for the elastic process, 
an infrared factor for the emission of a soft gluon and a noneikonal correction factor. The expression in (\ref{eq101}) will lead to $q_\bot \gg k_\bot$ limit of 
the Gunion and Bertsch formula~\cite{Gunion:1981qs}. 
If the emitted gluon is much softer than others it can then be 
regulated by the Debye screening mass, $m_d$. Terms within the parenthesis in (\ref{eq101})
would correspond to  noneikonal correction  over the eikonal Gunion Bertsch  
formula.  Eq. (\ref{eq101}) is complete upto ${\cal O}(1/k_\bot^2)$ and ${\cal O}(t/s)$, 
for emission of a soft gluon in the process $qq'\rightarrow qq'g$. 
Similar investigation have been done earlier for the process $gg \rightarrow ggg$ \cite{Das:2010hs,Abir:2010kc,Bhattacharyya:2011vy}.

\begin{figure}[t]
\includegraphics[width=3in,height=3in]{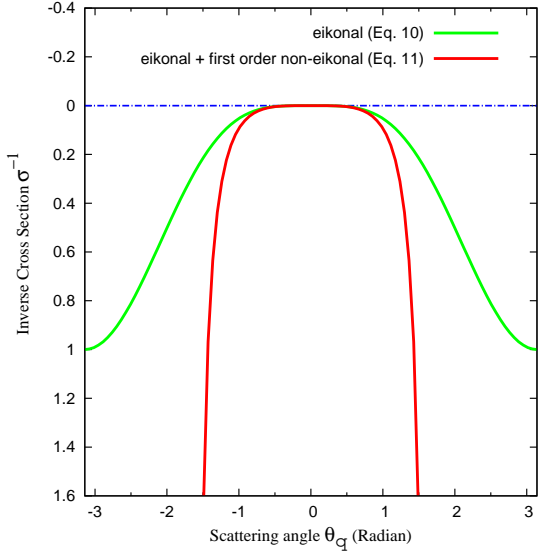}
\vspace*{-0.0in}
\caption{Dependence of inverse cross-section ${\sigma^{-1}}$ on scattering angle $\theta_q$, all prefactor that are independent of $\theta_q$ taken to be unity for convenience.}
\label{fig0}
\end{figure}

\subsubsection{Rutherford scattering beyond eikonal approximation} 
It is interesting to note how noneikonality gives way to probe beyond Rutherford scattering limits.
In the centre of momentum frame for a typical $2\rightarrow 2$ (or even in case of $2\rightarrow 3$ when fifth particle is ultra soft!) process
\begin{eqnarray}
\frac{t}{s}&=&-\sin^2\frac{\theta_q}{2}  \label{eq90}  
\end{eqnarray}
The eikonal cross-section $(\sigma_{eknl})$ is directly connected to the {\it Rutherford scattering} scattering cross-section as
\begin{eqnarray}
\sigma_{eknl} \propto \frac{s^2}{t^2} = \frac{1}{\sin^4(\theta_q/2)}\, .
\end{eqnarray}
When one relaxes the eikonal approximation then the cross-ection can be written as 
\begin{eqnarray}
\!\!\! \sigma_{ne}\!\! &\propto &\!\! \frac{s^2}{t^2}\left(1+\frac{17}{9}\frac{t}{s}\right) 
\!\!=\!\! \frac{1}{\sin^4\frac{\theta_q}{2}}\left(1-\frac{17}{9}\sin^2\frac{\theta_q}{2}\right),
\end{eqnarray}
which puts a restriction on the scattering angle, $\theta_q$. In Fig.\ref{fig0} we have plotted inverse cross-section $(\sigma^{-1})$  for both eikonal and 
noneikonal case. Even though both behave identically with a similar plateau in the small angle scattering but noneikonal cross-section 
has a very sharp fall in comparison with the eikonal one for large angle scattering. As seen the noneikonal inverse matrix element 
is bounded by the scattering angle, $\theta_q=\pm2\sin^{-1}(3/\sqrt{17})\simeq \pm 0.52 \pi$, in centre of momentum frame, in contrary to 
that of eikonal one having a natural bound of $\pm \pi$. This indicates that the back scattering is forbidden for the case of $qq'\rightarrow qq'g$ when the emitted gluon is soft.

\subsubsection{Cross-section in the first order in eikonal ({\it viz.},noneikonal) approximation}
The cross-section for the process ${qq'\rightarrow qq'g}$  can be obtained as  

\begin{widetext}
\begin{eqnarray}
\sigma_{qq'\rightarrow qq'g} &=& \frac{1}{2s} \int \prod_{i=3}^{5} \frac{d^3k_i}{(2\pi)^32E_i}
\left|{\cal M}_{qq'\rightarrow qq'g} \right| (2\pi)^4 \delta^{4}(k_1+k_2-k_3+k_4+k_5) \, .   
\end{eqnarray}
In the centre of momentum frame, ${\bf k_1}+{\bf k_2}={\bf k_3}+{\bf k_4}+{\bf k_5} = {\bf 0}$, and one obtains
\begin{eqnarray}
\sigma_{qq'\rightarrow qq'g} &=& \frac{1}{2s} \int \frac{d^3k_3}{(2\pi)^32E_3}
\frac{1}{(2\pi)^3 2E_4}\frac{d^3k_5}{(2\pi)^32 E_5} \left|{\cal M}_{qq'\rightarrow qq'g} \right| 
(2\pi)^4 \delta(E_1+E_2-E_3+E_4+E_5) \nonumber \\
&=& \frac{1}{2s} \left[-\frac{1}{2} \frac{1}{(2\pi)^2}\int \frac{dq_\bot^2 d q_{z}}{E_3}
\right]\frac{1}{(2\pi)^32E_4}
\left[\frac{1}{4} \frac{1}{(2\pi)^2}\int \frac{dk_\bot^2 d\theta_g}{\sin\theta_g}\right]\nonumber \\
&&~~~~~~~~~~~~~~~~~~~~ \times 12g^2\frac{8}{9}g^4\frac{s^2}{t^2}\frac{1}{k_\bot^2}\left(1+\frac{17}{9}
\frac{t}{s}\right)(2\pi)^4 \delta(E_1+E_2-E_3+E_4+\omega)  \nonumber \\
&=& \frac{1}{2s} \left[-\frac{1}{2} \frac{1}{(2\pi)^2}\int \frac{dq_\bot^2}{E_1}\right]
\frac{1}{(2\pi)^32E_1}\left[-\frac{1}{4} \frac{1}{(2\pi)^2}\int dk_\bot^2 d\eta\right]\nonumber \\
&&~~~~~~~~~~~~~~~~~~~~  \times 12g^2\frac{8}{9}g^4\frac{s^2}{(q_\bot^2)^2}\left(1+\frac{q_\bot^2}{s}\right)^{-2}\frac{1}{k_\bot^2}
\left(1-\frac{17}{9}\frac{q_\bot^2}{s}\right)(2\pi)^4 \, ,
\end{eqnarray}
\end{widetext}
where we have used rapidity $\eta = - \ln \left[\tan\left(\theta_g/2\right)\right]$, 
$\bf q = k_1-k_3$, $\bf q_\bot = q~{\sin \theta_q}$. The cross-section contains factors, having term like $q_\bot^2/s$,
that are responsible for non-eikonal effects.

In thermal medium taking debye mass as infrared regulator, the differential cross-section can be expressed as
\begin{eqnarray}
\frac{d~\sigma^{qq'\rightarrow qq'g}}{dq_\bot^2 dk_\bot^2 d\eta} = 2C_A C_{qq'} \alpha^3  \frac{\Gamma_{a b}}{(q_\bot^2+m_d^2)^2} \frac{1}{k_\bot^2+m_d^2} \, ,
\label{100}
\end{eqnarray}
where $C_A=3$ and  $C_{qq'}=8/9$ are Casimir factors, and $\Gamma_{ab}=\zeta_{\it a}\zeta_{\it b}$, with various factors are,
explicitly, given as
\begin{eqnarray}
\zeta_{\it a}&=&\left(1+\frac{q_\bot^2}{s}\right)^{-2} \, ,  \nonumber \\
\zeta_{\it b}&=&\left(1-\frac{17}{9}\frac{q_\bot^2}{s}\right) \ .
\end{eqnarray}
The factor  
$\zeta_{\it a}$ comes from eikonal part of the matrix elements, 
and $\zeta_{\it b}$ is the noneikonal factor originated from  noneikonal part of matrix elements. 
The differential cross-section for the process $qq'\rightarrow qq'g$ as given in Eq.(\ref{100}) correctly reproduces the result 
of \cite{Biro:1993qt} in the limit $q_\bot^2 \gg k_\bot^2$ and in the eikonal limit, $\sqrt{s}, E \gg q_\bot^2$, as all the 
noneikonal factors, $viz.,$ $\zeta_{\it a}$ and $\zeta_{\it b}$ become identically unity and so as $\Gamma_{ab}$.

\section{Inelastic gluon-gluon fusion ($gg\rightarrow ggg$)} 
The three gluon production via gluon-gluon fusion  $gg \rightarrow ggg$ is 
extremely important in the context of  heavy-ion phenomenology.
For a sequence of events:  hot glue scenario of glasma field, 
thermal equilibration, gluon chemical equilibration in later time, parton matter viscosity,
radiative energy-loss of high energy partons jet propagating through thermalised QGP, 
this process plays  a crucial role. 
Matrix elements for the process $gg \rightarrow ggg$ have been computed up to ${\cal O}(t^3/s^3)$ in 
\cite{Abir:2010kc}. 
Considering ${\cal O}(t/s)$ result it is now straightforward to 
evaluate the differential cross-section for this process in first order in eikonal approximation as 
\begin{eqnarray}
\frac{d~\sigma^{gg\rightarrow ggg}}{dq_\bot^2 dk_\bot^2 d\eta} = 2C_A C_{gg} \alpha^3  \frac{\Gamma_{a b}}{(q_\bot^2+m_d^2)^2} \frac{1}{k_\bot^2+m_d^2} \,
\label{110}
\end{eqnarray}
where $C_{gg}=9/2$ and the factor,  $\Gamma_{ab}=\zeta_{a}\zeta_{b}$, with its various components
\begin{eqnarray}
\zeta_{\it a}&=&\left(1+\frac{q_\bot^2}{s}\right)^{-2} \, , \nonumber \\
\zeta_{\it b}&=&\left(1-\frac{1}{2}\frac{q_\bot^2}{s}\right) \ .
\end{eqnarray}
\begin{figure}[t]
\includegraphics[width=1.0\columnwidth]{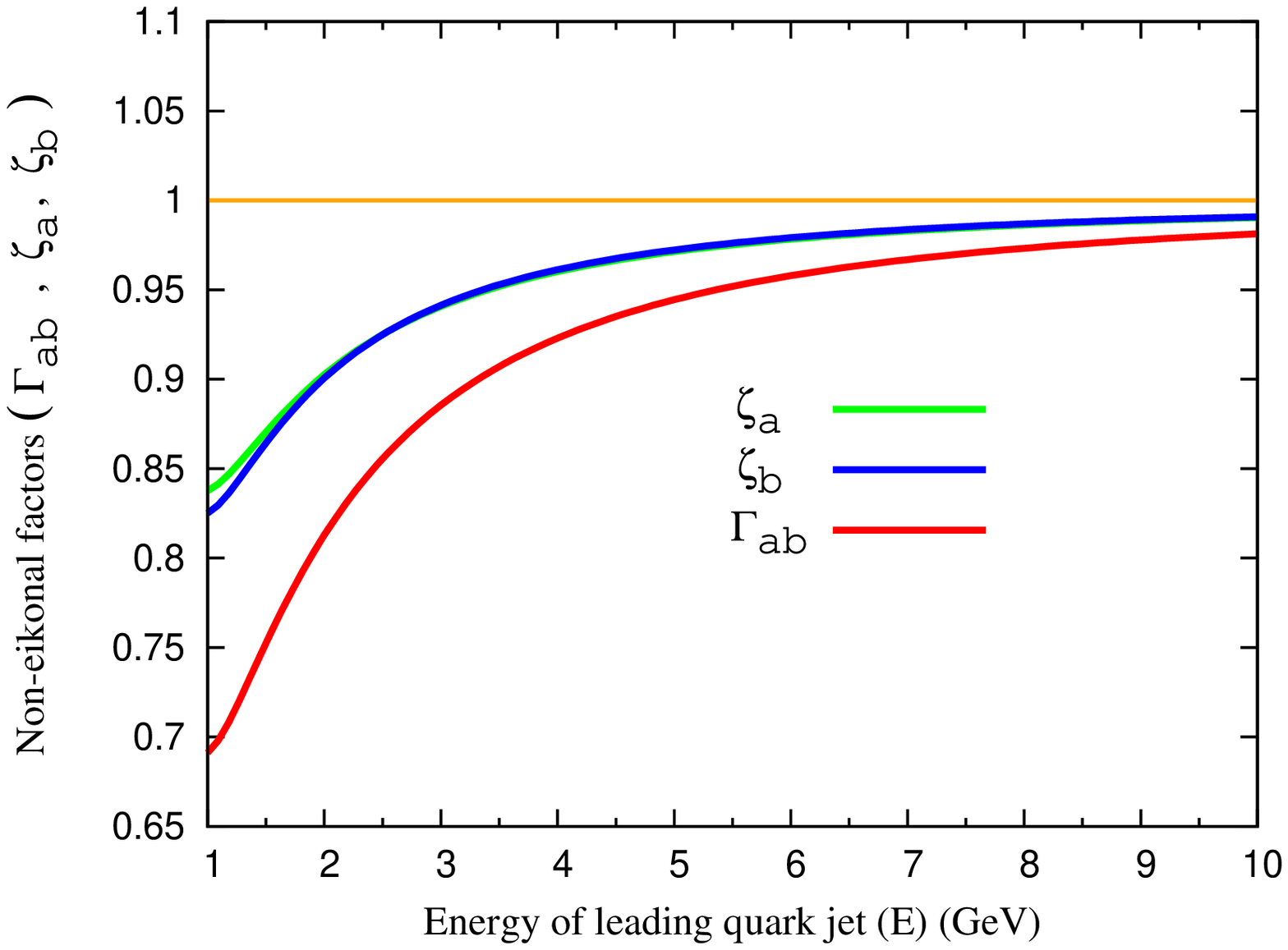}
\vspace*{-0.0in}
\caption{Typical estimation of noneikonal factors at $T= 300 MeV$ with $\alpha = 0.3$ for the process $qq'\rightarrow qq'g$.
First order noneikonal factors $\zeta_{a}=\left(1+{q_\bot^2}/{s}\right)^{-2}$, $\zeta_{b}=\left(1-{17q_\bot^2}/{9s}\right)$, and the full contribution ${\Gamma}_{ab} = \zeta_{a}\zeta_{b}$.}
\label{fig4}
\end{figure}

\begin{figure}[t]
\includegraphics[width=1.0\columnwidth]{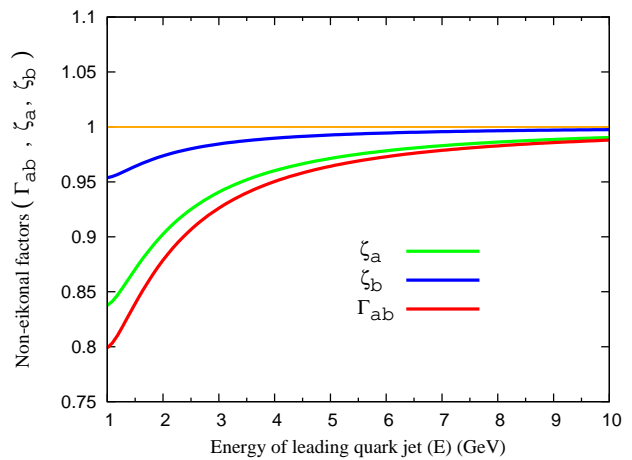}
\vspace*{-0.0in}
\caption{Typical estimation of noneikonal factors at $T= 300 MeV$ with $\alpha = 0.3$ for the process $gg\rightarrow ggg$.
First order noneikonal factors $\zeta_{a}=\left(1+{q_\bot^2}/{s}\right)^{-2}$, $\zeta_{b}=\left(1-{q_\bot^2}/{2s}\right)$ and the full contribution ${\Gamma}_{ab} = \zeta_{a}\zeta_{b}$.}
\label{fig5}
\end{figure} 
The factor comming from eikonal part of matrix element $\zeta_{\it a}$ is same for both processes $qq'\rightarrow qq'g$
and $gg\rightarrow ggg$ whereas the noneikonal factor $\zeta_{\it b}$ is different. This noneikonal factor, obviously, does not put any restriction on the scattering angle ($\theta=\pm\pi$) for the process $gg\rightarrow ggg$, and allows it to go in full natural range $\pm \pi$ as compared to the process $qq'\rightarrow qq'g$.

Unlike $gg\rightarrow ggg$ where a Park-Taylor type formula \cite{Parke:1986gb} is available to compute the matrix element, 
the computation of matrix elements up to ${\cal O}(t/s)$ is quite cumbersome in case of $qg\rightarrow qgg$. Also in this article we have 
performed our study on inelastic quark-quark scattering but with different flavours. In case of same flavour $qq\rightarrow qqg$ things would be a
more involved one. We leave them for future study.

\section{Results and Discussion} 
For quantitative estimation of the noneikonal effects, we have taken 
the average value of the momentum transfer squared which can be obtained \cite{Abir:2012pu} as
\begin{eqnarray}
\langle q_\bot^2\rangle &\simeq & \left (  {\int\limits_{m_g^2}^{E^2} dq_\bot^2 \ q_\bot^2 \ 
\frac{d\sigma_{2\rightarrow 3}}{dq_\bot^2}}\right ) / \left (
{\int\limits_{m_g^2}^{E^2} dq_\bot^2 \  
\frac{d\sigma_{2\rightarrow 3}}{dq_\bot^2}} \right )  \nonumber \\
&\simeq & 2 g^2T^2 \ln \left(E/gT\right).
\label{eq18}
\end{eqnarray}
$\langle q_\bot^2\rangle$ has then been embedded in $\zeta_{a}$, $\zeta_{b}$ to have a qualitative estimation of noneikonal effects over eikonal crosssections.

In Fig.\ref{fig4} and Fig.\ref{fig5} the first order noneikonal factors: $\zeta_{a}$, $\zeta_{b}$ and the full contribution $\Gamma_{ab}$ for both  processes $qq'\rightarrow qq'g$ and $gg\rightarrow ggg$, respectively,
displayed.
It can be seen that the noneikonal effects are  $\sim (15-20)\% $ over eikonal one for moderately hard jets. However, the noneikonal effect 
gradually becomes mild for very high energetic jets. Typically for charged hadrons in the momentum range of $8$ to $50$ GeV, i.e. typical parton kinematics of $16$ to $150$ GeV, way off the scale of Fig.\ref{fig4} and Fig.\ref{fig5}, indicate that
non-eikonal effects are largely absent above $(10-15)$ GeV parton kinematics. In the literature attempts have already been made to address the noneikonal 
propagation of partons for collision/elastic processes in a monte-carlo approach \cite{Auvinen:2009qm} by considering full
${\cal{O}}(\alpha_s^2)$ matrix elements for relevant $2\rightarrow 2$ processes. Eikonal propagation approximation 
was found to be good on the $10\%$ level. Present study also revels that for radiative/inelastic process, 
{\it Eikonal parton trajectory~I} approximation seems to be within $(15-20)\%$ level. 
This approximation should be crude only in soft and moderate momentum regimes. We also note that splitting kernels for partons produced in large virtuality scattering processes that subsequently traverse a region of strongly-interacting matter have been investigated early in the literature within effective theory formalism \cite{Ovanesyan:2011kn}.

There has always been a quest for large angle radiations. In the present study we do not assume small angle/collinear emission approximation 
either in the course of calculating matrix elements or in the calculations of kinematics.  
In the present calculation the kinematic relation $k_\bot = \omega \sin \theta_g$ ensures that one can go safely to the limit where $k_\bot \simeq \omega$.  

%

Most of the  pQCD based model calculations for describing the medium is not
able to account the almost ideal fluid behaviour, which seems to be a manifestation of long range correlations. 
It also would lead to a large elastic contribution to
energy loss \cite{Auvinen:2010yt} for reasonable values of coupling which is not supported by
the data. Sort of single hard scenario \cite{Gyulassy:2000fs,Wicks:2005gt} has been discussed here at the level of single emission karnel in which gluons are induced by a single hard scattering with the medium, but many
widely used quenching models, for instance \cite{Salgado:2003gb}, make the opposite assumption of multiple soft
interactions with the medium leading to induced gluon radiation. Such
classes of models can probably be expected to have somewhat milder
non-eikonal corrections due to scattering with the medium.

The kinematic constraints, $E \gg \omega \gg k_\bot,q_\bot$ referred in the literature as soft 
eikonal approximation that neglects any change in parton trajectory  due to multiple scatterings but assumes a straight line 
tragectory throughout. The diffusion of partons in a hot and dense medium can have an unavoidable link beyond the eikonal approximation and 
it is worth  to relax eikonal approximation.
In this work an attempt has been made to relax part of this approximation for some of the inelastic processes and their 
differential cross-sections in first order noneikonal approximation have been obtained. Primary estimation indicates $15-20\%$ reduction in 
the cross section due to first order noneikonal effect for both the processes in the soft and intermediate parton energies. These cross-sections  naturally reproduce eikonally approximated results in the eikonal limit for soft emission, {\it i.e.}, $\sqrt{s},E\gg q_\bot$ and $q_\bot \gg k_\bot$.
QGP produced at LHC, where large virtuality scattering processes may be dominant one, is seems to be $`$less opaque to jets than predicted' by constrained extrapolations from RHIC \cite{Horowitz:2011gd}. There are however other views also, for
instance \cite{Renk:2011aa,Renk:2011gj,Renk:2012wr}, where another set of constrained extrapolations 
show considerable variation in the postdictions
of RHIC-constrained scenarios with LHC data. Here
this have been taken as a constraint and
cause to disregard class of models which fail to predict/postdict
correctly the uprising behaviour of nuclear modification factor rather than assigning it as generic surprising feature of LHC data.

Our results indicate some reductions in interaction strengths of jets due to non-eikonal effects,
in soft and intermediate sector. In the soft sector when the problem is embedded 
into a hydrodynamically evolving density distribution this could lead to non-trivial effects.
We also show that wide back scattering with scattering angle more than $\simeq \pm 0.52 \pi$ is forbidden in case of $qq' \rightarrow qq'g$ when the emitted gluon in soft. 
This, however, is not the case for $gg\rightarrow ggg$.
In future study we intend  quantitative estimation of this noneikonal effect in jet quenching and 
other consequences in heavy-ion collisions phenomenology.

\vspace*{0.3in}

\begin{acknowledgments}
{\it Acknowledgments~:~}
I thank Munshi G. Mustafa for valuable discussions with numerous help during the course of this work and critically reading the manuscript. I also thank Jan Uphoff for his valuable suggestions and comments.

\end{acknowledgments}

\end{document}